\documentclass[12pt]{article}
\usepackage{epsfig}

\newcommand{\sll}{/\kern-4pt l}
\newcommand{\slp}{p\kern-5pt/}
\newcommand{\slq}{q\kern-5.5pt/}

\newcommand{\Disc}{{\rm Disc\,}}

\newcommand{\eps}{\varepsilon}
\newcommand{\pfrac}[2]{\left(\frac{#1}{#2}\right)} 
\newcommand{\simle}{\ \hbox{\raise2pt\rlap{$<$}%
  \lower3pt\rlap{$\sim$}\phantom{$<$}}\ }
\begin{document}
\thispagestyle{empty}
\begin{flushright}
MZ-TH/00-07\\
hep-ph/0003115\\
March 2000
\end{flushright}
\vspace{0.5cm}
\begin{center}
{\Large\bf 
Threshold expansion of Feynman diagrams\\
  within a configuration space technique}\\[1truecm]
{\large S.~Groote$^1$ and A.A.~Pivovarov$^{1,2}$}\\[.7cm]
$^1$ Institut f\"ur Physik, Johannes-Gutenberg-Universit\"at,\\[.2truecm]
  Staudinger Weg 7, D-55099 Mainz, Germany\\[.5truecm]
$^2$ Institute for Nuclear Research of the\\[.2truecm]
  Russian Academy of Sciences, Moscow 117312
\vspace{1truecm}
\end{center}

\begin{abstract}
The near threshold expansion of generalized sunset-type (water melon) diagrams
with arbitrary masses is constructed by using a configuration space technique.
We present analytical expressions for the expansion of the spectral density
near threshold and compare it with the exact expression obtained earlier using
the method of the Hankel transform. We formulate a generalized threshold
expansion with partial resummation of the small mass corrections for the
strongly asymmetric case where one particle in the intermediate state is much
lighter than the others.
\end{abstract}

\newpage

\section{Introduction}
Perturbation theory remains the most accurate and reliable tool of theoretical
analysis in elementary particle phenomenology. Even the most promising
approach based on the direct calculation of physical observables on the
lattice still cannot compete with perturbation theory supplemented with
some non-perturbative additions as condensates or renormalons. The accuracy of
experimental data for tests of the Standard Model permanently improves and 
demands the improvement of the accuracy of theoretical predictions~\cite{PDG}. 
Within perturbation theory this demand implies the calculation of Feynman
diagrams with an increasing number of loops (see e.g.~\cite{revloop}). At
present, Feynman diagrams of rather general form can be obtained analytically
at most up to three loops. Any of these calculations is still unique and
involves sophisticated algorithms and an extensive usage of computer resources
(see e.g.~\cite{ChKSt2,Broadhurst1}). Even the notion of the analytical
calculation is changing now because the analytical results are expressed
through rather special transcendental numbers which eventually have to be
calculated numerically. The striking example is the calculation of massive
three-loop bubbles~\cite{Broadhurst1} where the analytical results for the
diagrams are represented through the quantity which is given by a two-fold
infinite sum (see also~\cite{Osland}). Not all of the multiloop Feynman
diagrams are equally complicated. There are some topologies of $n$-loop
diagrams or special kinematic regimes for the external momenta which are much
simpler to compute than the general $n$-loop case (see e.g.~\cite{renZ}).
The increasing complication of loop calculations within perturbation theory is
more and more circumvented by using different kinds of expansion for the cases
where the kinematics allows one to find a small parameter. Heavy Quark
Effective Theory (HQET) and threshold expansions for heavy particle
production are recent examples for such a situation. These expansions are
mostly used for phenomenological applications where it even suffices to give
approximate numerical values for the unexpanded multiloop integrals. In some
cases one can obtain an analytical expansion that can be used for the
simplification of the calculations or for the evaluation of the exact
parameters of the effective Lagrangians through matching conditions. 

In this paper we discuss a method to calculate the near threshold expansion of
the spectral density for the class of multiloop sunset-type Feynman diagrams
termed water melon diagrams~\cite{GKP1}. These diagrams has recently drawn
some attention as a laboratory for testing some advanced methods in loop
calculations~\cite{Berends,Ussyukina}. However, water melon diagrams also have
numerous phenomenological applications. The most important application of
these diagrams (as well as spectacle diagrams as their first generalization)
is the calculation of the effective potential in quantum field theory both at
finite temperature and at zero
temperature~\cite{Jackiw,Rajantie,Chung,Braaten,Chung2,temp}. Among other
applications one can name the sum rule analysis of baryons in QCD both in the
massless approximation and with finite mass (the leading order correlator
being the standard sunset diagram) \cite{Ioffe,Kras,bar,barGro} or the
treatment of baryons in the large $N_c$ limit of QCD~\cite{NcHoo,NcBar} which
requires the calculation of $(N_c-1)$-loop water melon diagrams in the leading
order. An interesting application is the calculation of dibaryon properties in
operator product expansion within the sum rule approach which is important for
understanding many-quark bags in the nucleus~\cite{dibar,nucl}. The water
melon diagrams emerge in chiral perturbation theory~\cite{Post,Gasser}. The
corresponding integrals appear for the correlator of the effective operators
related to the mixing of neutral mesons in flavour dynamics~\cite{Narison} and
in the sum rule analysis of hybrid mesons~\cite{hyb1}. Water melon diagrams
constitute an important part of the contributions generated by the recurrence
relations of the integration-by-part techniques for three-loop
diagrams~\cite{Broadhurst2} (for the recent progress see e.g.\
Ref.~\cite{Baikov}). It is a rather hot topic and a dynamically developing
area of loop calculations where new results are frequently reported (see e.g.\
Refs.~\cite{Ligterink,Smirnov,KasteningKleinert}).

In our opinion, the method presented in Ref.~\cite{GKP1} completely solves
the problem of computing this class of diagrams. The method is simple and
reduces the calculation of a $n$-loop water melon diagram to a one-dimensional
integral which includes only well-known special functions in the integrand
(Bessel functions of different kinds) for any values of the internal masses.
The technique is universal. The most interesting part of our analysis of water
melon diagrams is the construction of the spectral decomposition of water
melon diagrams, i.e.\ the determination of the discontinuity across the
physical cut in the complex plane of the squared momentum. In this context a
novel technique for the direct construction of the spectral density of water
melon diagrams based on an integral transform in configuration space was
presented in Ref.~\cite{GKP1}. In the present paper we develop some explicit
expansions for water melon diagrams and compare them with the exact results.
The purpose of this paper is a three-fold one, namely:
\begin{itemize}
\item to demonstrate the ease with which threshold expansions to arbitrary
space-time dimensions and a general number of internal lines with arbitrary
masses can be generated within a configuration space technique. The entire
construction reduces to algebraic operations while the only integral
encountered is a simple integral of the type of Euler's Gamma function.
\item to generate explicit forms of threshold expansions and to analyze their
convergence properties. The explicit forms of threshold expansions for the
simple sunset can be compared with threshold expansions that are obtained in
momentum space.
\item to consider a case when one mass is much smaller than the others. This
is an important generalization of the threshold expansion and can be
analytically done within the configuration space technique.
\end{itemize}
The present analysis of the threshold expansion of water melon diagrams shows
the effectiveness of the configuration space technique for this topology class
of Feynman diagrams. On the other hand the obtained explicit results can be
useful for a variety of applications which include the evaluation of water
melon type diagrams. The technique can readily be generalized to some other
simple topologies of $n$-loop diagrams.

The paper is organized as follows. In Sec.~2 we provide the tools necessary
for the following calculations. Sec.~3 deals with the threshold expansion.
Here we outline our main strategy for this expansion in configuration space
and in the subsections that follow we give examples for the sunset diagram and
the water melon diagram with four or more internal lines. In Sec.~4 we
concentrate on the strongly asymmetric case where one of the masses is much
smaller than the others. We introduce the procedure of resummation of the
contributions of this smallest mass and show explicitly how we get to 
a closed expression for the spectral density even in this case. We give
examples for a strongly asymmetric mass arrangement for the water melon
diagrams with two (degenerate case), three, and four or more internal lines,
and compare the resummed results with the pure threshold expansion as well as
with the exact spectral density. In Sec.~5 we discuss how to recover the
non-analytic part of the polarization function through the spectral density
near threshold. In Sec.~6 we give our conclusions.

\section{Basics about the configuration space technique}
We start with a brief outline of the technique which we use in this
paper~\cite{GKP1}. The polarization function $\Pi(x)$ of a water melon diagram
including $n$ internal lines with masses $m_i$, $i\in\{1,\ldots,n\}$ in
configuration space is given by the product
\begin{equation}\label{prod}
\Pi(x)=\prod_{i=1}^nD(x,m_i).
\end{equation}
The propagator $D(x,m)$ of a massive line with mass $m$ in $D$-dimensional
(Euclidean) space-time is given by
\begin{equation}\label{prop}
D(x,m)=\frac1{(2\pi)^D}\int\frac{e^{ip_\mu x^\mu}d^Dp}{p^2+m^2}
  =\frac{(mx)^\lambda K_\lambda(mx)}{(2\pi)^{\lambda+1}x^{2\lambda}}
\end{equation}
where we write $D=2\lambda+2$. $K_\lambda(z)$ is a McDonald function
(a modified Bessel function of the third kind, see e.g.~\cite{Watson}).
In the limit $m\rightarrow 0$ the propagator in Eq.~(\ref{prop}) simplifies to
\begin{equation}
  D(x,0)=\frac1{(2\pi)^D}\int\frac{e^{ip_\mu x^\mu}d^Dp}{p^2}
  =\frac{\Gamma(\lambda)}{4\pi^{\lambda+1}x^{2\lambda}}
\end{equation}
where $\Gamma(\lambda)$ is Euler's Gamma function. Eq.~(\ref{prod}) contains
all information about the water melon diagrams and in this sense is the final
result for the class of diagrams under consideration. It is of course known
and was used since long ago~\cite{Mendels}. Of some particular interest is the
spectral decomposition of the polarization function $\Pi(x)$ which is
connected to the particle content of a given model. The spectrum of particles
is contained in the function $\rho(s)$ related to the polarization function
through the dispersion representation 
\begin{equation}\label{convo1}
\Pi(x)=\int_0^\infty\rho(s)D(x,\sqrt s)ds.
\end{equation}
The dispersion representation of the polarization function in configuration
space reveals the analytic structure of the polarization function $\Pi(x)$.
For applications, however, one may need the Fourier transform of the
polarization function $\Pi(x)$ given by
\begin{equation}\label{curr}
\Pi(q)=\int\Pi(x)e^{iq_\mu x^\mu}d^Dx. 
\end{equation}
(We use the same notation for the function and its Fourier transform because
we think that this will cause no confusion). Because of the Lorentz invariance
of the propagator the angular integration in Eq.~(\ref{curr}) can be done
explicitly in $D$-dimensional space-time (for a generalization to tensor
propagators see Ref.~\cite{GKP1}). The result reads
\begin{equation}\label{measure}
\int e^{iq_\mu x^\mu}d^D\Omega
  =2\pi^{\lambda+1}\left(\frac{|q||x|}2\right)^{-\lambda}J_\lambda(|q||x|)
\end{equation}
where $J_\lambda(z)$ is the usual Bessel function and $d^D\Omega$ is the
rotationally invariant measure on the unit sphere in the $D$-dimensional
(Euclidean) space-time. Note that the polarization function $\Pi(x)$ as well
as its Fourier transform $\Pi(q)$ are only functions of the absolute value
$|x|$ and $|q|$. The same is of course valid also for the other occurring
functions. To simplify the notation we often write $x=|x|$ and
$q=|q|$ for these absolute values.

Note that propagators of particles with non-zero spin in configuration space 
representation can be obtained from the scalar propagator by differentiation
with respect to the space-time point $x$. This does not change the functional
$x$-structure and causes only minor modification of the basic technique
(for details see Ref.~\cite{GKP1} and references therein). Our final
representation of the Fourier transform of a water melon diagram is given by
the one-dimensional integral
\begin{equation}\label{exactpi}
\Pi(q)=2\pi^{\lambda+1}\int_0^\infty\left(\frac{qx}2\right)^{-\lambda}
  J_\lambda(qx)\Pi(x)x^{2\lambda+1}dx
\end{equation}
which is a special kind of integral transformation with a Bessel function 
as a kernel. This integral transformation is known as the Hankel
transform~\cite{Meijer,Erdelyi}. The representation given by
Eq.~(\ref{exactpi}) is quite universal regardless of whether tensor structures
are added or particles with vanishing momenta are radiated from any of the
internal lines. Note in this context that Bessel functions are objects
well-studied during the last two centuries. They therefore can be added to the
list of elementary functions.

Because the dispersion representation of the polarization function in
configuration space (or the spectral density of the corresponding polarization
operator) has the form given in Eq.~(\ref{convo1}), the analytic structure of
the polarization function $\Pi(x)$ can be determined directly in configuration
space without having to compute its Fourier transform first. The
transformation in Eq.~(\ref{convo1}) turns out to be a particular example of
the Hankel transform, namely the $K$-transform~\cite{Meijer,Erdelyi}. As
pointed out in Ref.~\cite{GKP1}, the inverse $K$-transform in this case is
given by
\begin{equation}\label{exactrho}
\rho(s)=\frac{(2\pi)^\lambda}{is^{\lambda/2}}\int_{c-i\infty}^{c+i\infty}
  I_\lambda(\zeta\sqrt s)\Pi(\zeta)\zeta^{\lambda+1}d\zeta
\end{equation}
where $I_\lambda(z)$ is a modified Bessel function of the first kind and the
integration runs along a vertical contour in the complex plane to the right 
of the right-most singularity of $\Pi(\zeta)$. The inverse transform given by
Eq.~(\ref{exactrho}) completely solves the problem of determining the spectral
density of the general class of water melon diagrams with any number of
internal lines and different masses by reducing it to the computation of a
one-dimensional integral. For $n$ internal lines with equal masses $m$ the
spectral density reads explicitly
\begin{equation}
\rho(s)=\frac{m^{\lambda n}}{i(2\pi)^{(n-1)\lambda+n}s^{\lambda/2}}
  \int_{c-i\infty}^{c+i\infty}I_\lambda(\zeta\sqrt s)
  \left(K_\lambda(m\zeta)\right)^n\zeta^{1-(n-1)\lambda}d\zeta.
\end{equation}
In contrast to our technique, in the standard, or momentum, representation
the polarization function $\Pi(q)$ is calculated from a $(n-1)$-loop diagram
with $(n-1)$ $D$-dimensional integrations over the entangled loop momenta
which makes the computation difficult when the number of internal lines
becomes large.

\section{Threshold expansion}
With our method described in detail in Ref.~\cite{GKP1} the $s$-dependence of
the spectral density can be calculated by a one-fold numerical integration
according to Eq.~(\ref{exactrho}). The numerical integration in
Eq.~(\ref{exactrho}) can be done for arbitrary space-time dimensions and a
general number of lines with arbitrary masses. In this sense this is the most
efficient representation for the spectral density of the water melon diagram.
However, we can also develop an explicit expansion near the threshold with any
desired accuracy. It does not require any complicated integrations at all. The
corresponding expansion of Eq.~(\ref{exactrho}) can then be compared with
series expansions near the production threshold obtained with the traditional
momentum space technique. We stress that we are only interested in the
spectral density because it is the main object for physical applications (see
e.g.\ Refs.~\cite{Penin}). For practical reasons we start with
Eq.~(\ref{exactpi}). The polarization function $\Pi(q)$ as written in
Eq.~(\ref{exactpi}) is, in general, UV divergent. The divergence can be
subtracted by using the power series expansion of the weight function
$(qx/2)^{-\lambda}J_\lambda(qx)$ to an appropriate order which will be added
and subtracted to this weight. This leads to a $q^2$-dependent power series of
divergent subtraction terms plus an UV finite subtracted integral (see
e.g.~\cite{GKP1,BogS}). But because the subtraction terms do not contribute to
the spectral density, we can avoid this subtraction at all. In order that the
formally written expressions make sense they are supposed to be dimensionally
regularized. We use the simplified or unorthodox dimensional regularization
method for water melon diagrams (see Ref.~\cite{GKP1}).

The threshold region of a water melon diagram is determined by the condition
$q^2+M^2\simeq 0$ where $q$ is the Euclidean momentum and $M=\sum_im_i$ is the
threshold value for the spectral density. We introduce the Minkowskian
momentum $p$ defined by $p^2=-q^2$ which is an analytic continuation to the
physical cut. Operationally this analytic continuation can be performed by
replacing $q\rightarrow ip$. To analyze the region near the threshold we use
the parameter $\Delta=M-p$ which takes complex values. The parameter $\Delta$
is more convenient in Euclidean domain while the parameter $E=-\Delta=p-M$ is
the actual energy counted from threshold which is used in phenomenological
applications. The spectral density as a function of $E$ is written as
$\tilde\rho(E)=\rho((M+E)^2)$ in the following. The analytic continuation of
the Fourier transform in Eq.~(\ref{exactpi}) to the Minkowskian domain has the
form
\begin{equation}
\Pi(p)=2\pi^{\lambda+1}\int_0^\infty\pfrac{ipx}2^{-\lambda}
  J_\lambda(ipx)\Pi(x)x^{2\lambda+1}dx.
\end{equation}
For the threshold expansion we have to analyze the large $x$ behaviour of the
integrand. It is this region that saturates the integral in the limit
$p\rightarrow M$ or, equivalently, $E\rightarrow 0$. It is convenient to
perform the analysis in a basis where the integrand has a simple large $x$
behaviour. The most important part of the integrand is the Bessel function
$J_\lambda(ipx)$ which, however, contains both rising and falling branches at
large $x$. It resembles the situation with elementary trigonometric functions 
$\sin(z)$ and $\cos(z)$ to which the Bessel function $J_\lambda(z)$ is rather
close (in a certain sense). Indeed, $\cos(z)$ (or $\sin(z)$) is a linear
combination of exponentials, namely
\begin{equation}
\cos(z)=\frac12\left(e^{iz}+e^{-iz}\right)
\end{equation}
and has also both rising and falling branches at large pure imaginary
argument: the exponentials show simple asymptotic behaviour $e^{\pm z}$ at
$z=\pm i\infty$. The analogous statement is true for $J_\lambda(z)$ which can
be written as a sum of two Hankel functions,
\begin{equation}\label{hansum}
J_\lambda(z)=\frac12(H_\lambda^+(z)+H_\lambda^-(z))
\end{equation}
where $H_\lambda^\pm(z)=J_\lambda(z)\pm iY_\lambda(z)$. The Hankel functions
$H_\lambda^\pm(z)$ show simple asymptotic behaviour at infinity,
\begin{equation}
H_\lambda^\pm(iz)\sim z^{-1/2}e^{\pm z}.
\end{equation}
Accordingly we split up $\Pi(p)$ into $\Pi(p)=\Pi^+(p)+\Pi^-(p)$ with
\begin{equation}
\label{pm}
\Pi^\pm(p)=\pi^{\lambda+1}\int_0^\infty\pfrac{ipx}2^{-\lambda}
  H_\lambda^\pm(ipx)\Pi(x)x^{2\lambda+1}dx.
\end{equation}
The two parts $\Pi^\pm(p)$ of the polarization function $\Pi(p)$ have
completely different behaviour near threshold which allows one to analyze them
independently. This observation makes the subsequent analysis straightforward.
We first consider the contribution of the part $\Pi^+(p)$. The behaviour at
large $x$ is given by the asymptotic form of the functions which we simply
write up to the leading terms as
\begin{equation}
H^+(ipx)=\sqrt{\frac{2}{i\pi px}}e^{-px}(1+O(x^{-1})),\qquad
  K(mx)=\sqrt{\frac{\pi}{2 m x}}e^{-mx}(1+O(x^{-1})).
\end{equation}
The large $x$ range of the integral (above a reasonably large cutoff parameter
$\Lambda$) has the general form
\begin{equation}\label{pilap}
\Pi^+_\Lambda(M-\Delta)\sim\int_\Lambda^\infty x^{-a}e^{-(2M-\Delta)x}dx
\end{equation}
where
\begin{equation}\label{defa}
a=(n-1)(\lambda+1/2).
\end{equation}
The right hand side of Eq.~(\ref{pilap}) is an analytic function in $\Delta$
in the vicinity of $\Delta=0$. It exhibits no cut or other singularities near
the threshold and therefore does not contribute to the spectral density. We
turn now to the second part $\Pi^-(p)$. In contrast to the previous case, the
integrand of this part contains $H^-(ipx)$ which behaves like a rising
exponential function at large $x$,
\begin{equation}
H^-(ipx)\sim x^{-1/2}e^{px}.
\end{equation}
Therefore the integral is represented by
\begin{equation}\label{pilam}
\Pi^-_\Lambda(M-\Delta)\sim\int_\Lambda^\infty x^{-a}e^{-\Delta x}dx.
\end{equation}
The function $\Pi^-(M-\Delta)$ is non-analytic near $\Delta=0$ because for
$\Delta<0$ the integrand in Eq.~(\ref{pilam}) grows in the large $x$ region
and the integral diverges at the upper limit. Therefore the function which is
determined by this integral is singular at $\Delta<0$ ($E>0$) and requires
an interpretation for these values of the argument $\Delta$. In the complex
$\Delta$ plane with a cut along the negative axis the function is analytic.
This cut corresponds to the physical positive energy cut. The discontinuity
across the cut gives rise to the non-vanishing spectral density of the
diagram.

\begin{figure}[th]
\begin{center}
\epsfig{figure=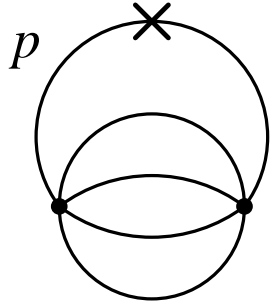, height=4truecm, width=3.5truecm}
\end{center}
\caption{\label{fig1}Representation of $\Pi^+(p)$ as vacuum bubble with added
line. The cross denotes an arbitrary number of derivatives to the specified
line.}
\end{figure}

Let us first discuss the analytic part $\Pi^+(p)$ of the diagram. This part
reduces to a regular water melon diagram. Indeed, using the relation
\begin{equation}\label{KtoH}
K_\lambda(z)=\frac{\pi i}2e^{i\lambda\pi/2}H_\lambda^+(iz)
\end{equation}
between Bessel functions of different kinds one can replace the Hankel
function $H_\lambda^+(ipx)$ with the McDonald function $K_\lambda(px)$. Since 
the propagator of a massive particle (massive line in the diagram) is given by
the McDonald function up to a power in $x$, this substitution shows that the
weight function behaves like a propagator of an additional line with the
``mass'' $p$. The explicit expression is given by
\begin{equation}
\Pi^+(p)=\frac{(-2\pi i)^{2\lambda+1}}{(p^2)^\lambda}
  \int_0^\infty\Pi_+(x)x^{2\lambda+1}dx.
\end{equation}
$\Pi_+(x)=\Pi(x)D(x,p)$ is the polarization function of a new effective
diagram which is equal to the initial polarization function multiplied by a
propagator with $p$ as mass parameter. We thus end up with a vacuum bubble of
the water melon type with one additional line compared to the initial diagram
(see Fig.~\ref{fig1}).
These diagrams have no singular behaviour at the production threshold $p=M$.
As mentioned above, $\Pi^+(p)$ is analytic in $\Delta$ near the origin
$\Delta=0$ and can therefore be omitted in the calculation of the spectral
density. All derivatives of $\Pi^+(p)\equiv\Pi^+(M-\Delta)$ with respect to
$\Delta$ are represented as vacuum bubbles with one additional line carrying
rising indices. Such diagrams can be efficiently calculated within the
recurrence relation technique developed in Ref.~\cite{GKP1}.   

Therefore one is left with the second part $\Pi^-(p)$ containing the
non-analytical contributions in $\Delta$ which leads to a non-vanishing
spectral density. Still the integral in Eq.~(\ref{pm}) cannot be done
analytically. In order to obtain an expansion for the spectral density near
the threshold in an analytical form we make use of the asymptotic series
expansion for the function $\Pi(x)$ which crucially simplifies the integrands
but still preserves the singular structure of the integral in terms of the
variable $\Delta$. The asymptotic series expansion to  order $N$ of the main
part of each propagator, i.e.\ of the McDonald function, is given
by~\cite{Watson}
\begin{equation}\label{Kas}
K_{\lambda,N}^{as}(z)=\pfrac\pi{2z}^{1/2}e^{-z}\Bigg[\sum_{n=0}^{N-1}
  \frac{(\lambda,n)}{(2z)^n}+\theta\frac{(\lambda,N)}{(2z)^N}\Bigg],\quad
  (\lambda,n):=\frac{\Gamma(\lambda+n-1/2)}{n!\Gamma(\lambda-n-1/2)}
\end{equation}
($\theta\in[0,1]$). Therefore the asymptotic expansion of the function
$\Pi(x)$ consists of an exponential factor $e^{-Mx}$ and an inverse power
series in $x$ up to an order $\tilde N$ which is closely related to $N$. It is
this asymptotic expansion that determines the singularity structure of the
integral. We write the whole integral in the form of the sum of two terms,
\begin{eqnarray}\label{pidias}
\Pi^-(p)&=&\pi^{\lambda+1}\int\pfrac{ipx}2^{-\lambda}H_\lambda^-(ipx)
  \left(\Pi(x)-\Pi^{as}_N(x)\right)x^{2\lambda+1+2\eps}dx\\&&
  +\pi^{\lambda+1}\int\pfrac{ipx}2^{-\lambda}H_\lambda^-(ipx)
  \Pi^{as}_N(x)x^{2\lambda+1+2\eps}dx=\Pi^{di}(p)+\Pi^{as}(p).\nonumber
\end{eqnarray}
The integrand of the first term $\Pi^{di}(p)$ behaves as $1/x^{\tilde N}$ at
large $x$ while the integrand of the second term accumulates all lower powers
of the large $x$ expansion. Note that only the large $x$ behaviour is
essential for the near threshold expansion of the spectral density. This fact
has been taken into account in Eqs.~(\ref{pilap}) and~(\ref{pilam}) where we
introduced a cutoff $\Lambda$. However, from the practical point of view the
calculation of the regularized integrals with an explicit cutoff is
inconvenient. The final result of the calculation -- the spectral density of
the diagram -- is independent of the cutoff, but the integration is
technically complicated if the cutoff is introduced. However, in extending the
integration over the whole region of the variable $x$ without using the cutoff
one immediately encounters divergences at small $x$ because the asymptotic
expansion is invalid in the region near the origin, so one is not allowed to
continue the integration to this region. The standard way to cope with such a
situation is to introduce dimensional regularization. It allows one to deal
with divergent expressions at intermediate stages of the calculation and is
technically simple because it does not introduce any cutoff and therefore does
not modify the integration region drastically. Note that dimensional
regularization does not necessarily regularize all divergences in this case
(in contrast to the standard case of ultraviolet divergences) but
nevertheless suffices for our purposes. We use a parameter $\eps$ to
regularize the divergences at small $x$. This regularization prescription is
an unorthodox version of the dimensional regularization (see e.g.\
Refs.~\cite{Pivovarov}).

The first part $\Pi^{di}(p)$ in Eq.~(\ref{pidias}) containing a difference 
of the polarization function and its asymptotic expansion as the integrand
gives no contributions to the spectral density up to a given order of the
expansion in $\Delta$. This is because the subtracted asymptotic series to
order $N$ cancels the inverse power behaviour of the integrand to this degree
$N$. The integrand decreases sufficiently fast for large values of $x$ and the
integral converges even at $\Delta=0$. Therefore this term is inessential when
the expansion of the spectral density is evaluated up to some order. One can
readily determine the order of the expansion near $\Delta=0$ at which a
contribution to the spectral density appears in using some further
simplifications of the integrand of the term $\Pi^{di}(p)$ in
Eq.~(\ref{pidias}). Namely, we replace the Hankel function under the
integration sign by its asymptotic series expansion. The resulting exponential
factor $e^{(p-M)t}$ can then be expanded in the parameter $\Delta=M-p$ and
integrated together with the finite inverse power series in $x$. One obtains a
finite power series in this parameter $\Delta$ which leads to a non-regular
term of order $\Delta^N$ (for instance, $\Delta^N\ln\Delta$ or
$\Delta^N\sqrt\Delta$). Therefore the part $\Pi^{di}(p)$ is regular and gives
no contribution to the spectral density up to the order $\Delta^N$. For this
reason we concentrate on the expansion of the second part $\Pi^{as}(p)$ and
find that only this part contains the contribution to the spectral density up
to the $N$.

Therefore the expansion of the spectral density at small $E$ is determined
only by the integral $\Pi^{as}(p)$ of Eq.~(\ref{pidias}). This integral is
still rather complicated to compute but we can go a step further in its
analytical evaluation. Indeed, since the singular behaviour of $\Pi^{as}(p)$
is determined by the behaviour at large $x$, we can replace the first factor,
i.e.\ the Hankel function, in the large $x$ region by its asymptotic expansion 
up to some order $N$. We use
\begin{equation}\label{Has}
H_{\lambda,N}^{-{as}}(iz)=\pfrac2{\pi z}^{1/2}e^{z+i\lambda\pi/2}
  \left[\sum_{n=0}^{N-1}\frac{(-1)^n(\lambda,n)}{(2z)^n}
  +\theta\frac{(-1)^N(\lambda,N)}{(2z)^N}\right]
\end{equation}
(cf.\ Eq.~(\ref{Kas}) for the notation) to obtain a representation
\begin{equation}\label{pidas}
\Pi^{das}(p)=\pi^{\lambda+1}\int\pfrac{ipx}2^{-\lambda}
  H_{\lambda,N}^{-as}(ipx)\Pi^{as}_N(x)x^{2\lambda+1+2\eps}dx.
\end{equation}
The index ``{\it das}'' stands for ``double asymptotic'' and indicates that
the integrand in Eq.~(\ref{pidas}) consists of a product of two asymptotic
expansions: one for the polarization function $\Pi(x)$ and another for the
Hankel function $H_{\lambda}^-(x)$ as weight (or kernel). Both asymptotic
expansions are straightforward and can be obtained from standard handbooks on
Bessel functions. We therefore arrive at our final result: the integration
necessary for evaluating the near threshold expansion of the water melon
diagrams reduces to integrals of the type of Euler's Gamma function, i.e.\
integrals containing exponentials and powers. Indeed, the result of the
expansion in Eq.~(\ref{pidas}) is an exponential function $e^{-\Delta x}$
times a power series in $1/x$, namely
\begin{equation}\label{serx}
x^{-a+2\eps}e^{-\Delta x}\sum_{j=0}^{N-1}\frac{A_j}{x^j}
\end{equation}
where $a$ has already been defined in Eq.~(\ref{defa}) and the coefficients
$A_j$ are simple functions of the momentum $p$ and the masses $m_i$. The
expression in Eq.~(\ref{serx}) can be integrated analytically using
\begin{equation}
\int_0^\infty x^{-a+2\eps}e^{-\Delta x}dx=\Gamma(1-a+2\eps)\Delta^{a-1-2\eps}.
\end{equation}
The result is
\begin{equation}\label{serint}
\Pi^{das}(M-\Delta)=\sum_{j=0}^{N-1}A_j\Gamma(1-a-j+2\eps)\Delta^{a+j-1-2\eps}.
\end{equation}
This expression is our final representation for the part of the polarization
function of a water melon diagram necessary for the calculation of the
spectral density near the production threshold. It is also one of the main
results of our paper.

Next we discuss the general structure of the expression in Eq.~(\ref{serint})
in detail. In the case where $a$ takes integer values, these coefficients
result in $1/\eps$-divergences for small values of $\eps$. The powers of
$\Delta$ in Eq.~(\ref{serint}) have to be expanded to first order in $\eps$
and give
\begin{equation}
\frac1{2\eps}\Delta^{2\eps}=\frac1{2\eps}+\ln\Delta+O(\eps).
\end{equation}
Because of
\begin{equation}
{\rm Disc\,}\ln(\Delta) \equiv \ln(-E-i0)-\ln(-E+i0)
=-2\pi i\theta(E)
\end{equation}
$\Pi^{das}(M-\Delta)$ in Eq.~(\ref{serint}) contributes to the spectral
density. For half-integer values of $a$ the power of $\Delta$ itself has a
cut even for $\eps=0$. The discontinuity is then given by
\begin{equation}
{\rm Disc\,}\sqrt{\Delta}=-2i\sqrt{E}\,\theta(E).
\end{equation}
Our method to construct a threshold expansion thus simply reduces to the
analytical calculation of the integral in Eq.~(\ref{pidas}) which can be done
for arbitrary dimension and an arbitrary number of lines with different
masses. In the next subsections we use our technique to work out some specific
examples which demonstrate both the simplicity and efficiency of our method.

\subsection{Equal mass sunset diagram}
The polarization function represented by the sunset diagram with three
propagators with equal masses $m$ in $D=4$ space-time dimensions is given by
\begin{equation}
\Pi(x)=\frac{m^3 K_1(mx)^3}{(2\pi)^6x^3}.
\end{equation}
The exact spectral density
is given by the integral representation in Eq.~(\ref{exactrho}) which for this
particular case reads
\begin{equation}\label{exactrho4}
\rho(s)=\frac{2\pi}{i\sqrt s}\int_{c-i\infty}^{c+i\infty}
  I_1(\zeta\sqrt s)\Pi(\zeta)\zeta^2d\zeta.
\end{equation}
In order to obtain a threshold expansion of the spectral density in
Eq.~(\ref{exactrho4}) we use Eq.~(\ref{serint}) to calculate the expansion of
the appropriate part of the polarization function. To illustrate the procedure
we present the explicit shape of the integrand in Eq.~(\ref{pidas}) which is
given by an asymptotic expansion at large $x$,
\begin{eqnarray}\label{ex}
\lefteqn{\pi^2\pfrac{ipx}2^{-1}H_{1,N}^{as}(px)\Pi_N^{as}(x)
  x^{3+2\eps}\ =\ \frac{m^{3/2}e^{(p-3m)x}}{(4\pi)^3p^{3/2}}x^{-3+2\eps}\
  \times}\nonumber\\&&\qquad\times\ \left\{1+\frac9{8mx}-\frac3{8px}
  +\frac9{128m^2x^2}-\frac{27}{64mpx^2}-\frac{15}{128p^2x^2}+O(x^{-3})\right\}.
\end{eqnarray}
From Eq.~(\ref{ex}) we can easily read off the coefficients $A_j$ that enter
the expansion in Eq.~(\ref{serx}). The spectral density is obtained by
performing the term-by-term integration of the series in Eq.~(\ref{ex}) and
by evaluating the discontinuity across the cut along the positive energy axis
$E>0$. The result reads
\begin{eqnarray}
\label{pidas430}
\lefteqn{\tilde\rho(E)\ =\ \frac{E^2}{384\pi^3\sqrt 3}
  \Bigg\{1-\frac12\eta+\frac7{16}\eta^2-\frac38\eta^3
  +\frac{39}{128}\eta^4-\frac{57}{256}\eta^5}\qquad\\&&
  +\frac{129}{1024}\eta^6-\frac3{256}\eta^7
  -\frac{4047}{32768}\eta^8+\frac{18603}{65536}\eta^9
  -\frac{248829}{524288}\eta^{10}+O(\eta^{11})\Bigg\}\nonumber
\end{eqnarray}
where the notation $\eta=E/M$, $M=3m$ is used. The simplicity of the
derivation is striking. By no cost it can be generalized to any number of
lines, arbitrary masses, and any space-time dimension. The standard equal mass
sunset is chosen for the definiteness only. It also allows us to compare our
results with results available in the literature. Eq.~(\ref{pidas430})
reproduces the expansion coefficients $\tilde a_j$ obtained in
Ref.~\cite{Smirnov} (the fourth column in Table~1 of Ref.~\cite{Smirnov}) by a
direct integration in momentum space within the technique of region
separation~\cite{Beneke}.

The case of the equal mass standard sunset diagram is the simplest one. There
exists an analytical expression for the spectral density of the sunset diagram
with three equal mass propagators in $D=4$ space-time dimensions in terms of
elliptic integrals~\cite{Prudnikov} (see also
Ref~\cite{ellRef}\footnote{We thank A.~Davydychev for bringing these papers 
to our attention.}). This expression can be used for a comparison with our
exact result in Eq.~(\ref{exactrho4}) or with the expansion in
Eq.~(\ref{pidas430}). However, we only present the result for $D=2$ in order
to keep the resulting expressions in a reasonably short form (cf.\
Ref.~\cite{GKP2}). In $D=2$ space-time dimensions the spectral density for a
sunset diagram with equal masses $m$ can be readily obtained. We just use the
exact expression for the spectral density in the convolution
representation~\cite{GKP1} and proceed towards $n=3$ equal masses. The
convolution function for two spectral densities in $D=2$ dimensional
space-time ($\lambda=0$) reads
\begin{equation}\label{convolut}
\rho(s;s_1;s_2)=\frac1{2\pi\sqrt{(s-s_1-s_2)^2-4s_1s_2}}.
\end{equation}
The two spectral densities one has to convolute are the spectral density of a
correlator with two equal masses and the spectral density of a single massive
line. While the latter is given by $\rho(s;m^2)=\delta(s-m^2)$, the former can
be obtained from Eq.~(\ref{convolut}) by inserting $s_1=s_2=m^2$,
\begin{equation}
\rho(s;m^2;m^2)=\frac1{2\pi\sqrt{s(s-4m^2)}}.
\end{equation}
So the convolution leads to
\begin{eqnarray}\label{ellint}
\lefteqn{\rho(s;m^2;m^2;m^2)\ =\ \frac1{4\pi^2}\int_{4m^2}^{(\sqrt s-m)^2}
  \frac{dt}{\sqrt{(s-m^2-t)^2-4m^2t}\sqrt{t(t-4m^2)}}}\nonumber\\
  &=&\frac1{4\pi^2}\int_{4m^2}^{(\sqrt s-m)^2}\frac{dt}{\sqrt{t(t-4m^2)
  ((\sqrt s+m)^2-t)((\sqrt s-m)^2-t)}}.
\end{eqnarray}
Now we use the relation (cf.\ Ref.~\cite{Prudnikov})
\begin{equation}\label{elldef}
\int_{t_1}^{t_2}\frac{dt}{\sqrt{(t-t_0)(t-t_1)(t_2-t)(t_3-t)}}
  =\frac2{\sqrt{(t_3-t_1)(t_2-t_0)}}K(k^2),
\end{equation}
\begin{equation}
k^2=\frac{(t_2-t_1)(t_3-t_0)}{(t_3-t_1)(t_2-t_0)}
\end{equation}
with $t_3>t_2>t>t_1>t_0$ and the definition of the complete elliptic integral
of the first kind
\begin{equation}
K(k^2)=\int_0^{\pi/2}\frac{d\varphi}{\sqrt{1-k^2\sin^2\varphi}}
  =F\left(\frac\pi2,k^2\right)
\end{equation}
(remark the difference in the definition) for $t_0=0$, $t_1=4m^2$,
$t_2=(\sqrt s-m)^2$, and $t_3=(\sqrt s+m)^2$ to perform the integration in
Eq.~(\ref{ellint}). We obtain
\begin{equation}
k^2=\frac{((\sqrt s-m)^2-4m^2)(\sqrt s+m)^2}
{((\sqrt s+m)^2-4m^2)(\sqrt s-m)^2}
\end{equation}
and finally end up with
\begin{equation}
\rho(s;m^2;m^2;m^2)=\frac{K(k^2)}{2\pi^2(\sqrt s-m)\sqrt{(\sqrt s+m)^2-4m^2}}.
\end{equation}
Therefore the spectral density in terms of the energy $E$ reads (see e.g.\
Ref.~\cite{Tokarev})
\begin{eqnarray}\label{ell}
\tilde\rho(E)&=&\frac1{2\pi^2(2m+E)\sqrt{(4m+E)^2-4m^2}}K(k^2),\nonumber\\
k^2&=&\frac{((2m+E)^2-4m^2)(4m+E)^2}{((4m+E)^2-4m^2)(2m+E)^2},\qquad M=3m.
\end{eqnarray}
By expanding the elliptic integral in terms of the threshold parameter $E$ 
one reproduces the threshold expansion in Eq.~(\ref{pidas430}). The result for
$D=4$ space-time dimensions is expressible by the elliptic integrals with
some rational functions as factors that makes the result a bit longer. Note
that the representation in Eq.~(\ref{ell}) is understood to be an analytical
expression for the spectral density. However, it is a matter of taste whether
the representation through the elliptic integrals as in Eq.~(\ref{ell}) is
considered simpler (or in a more analytical form) than the integral
representation in Eq.~(\ref{exactrho}). The only objection against the latter
which one can find in the literature is that the Bessel functions are
complicated (see e.g.\ Ref.~\cite{Ligterink}). But after more than a century
of intensive investigation they are well-known and no more complicated than
the square root of the fourth order polynomial which is used in
Eq.~(\ref{elldef}) to define the elliptic integral.

\subsection{Equal mass water melon diagrams\\ with four or more propagators}
The water melon diagrams with four or more propagators cannot be easily done 
by using the momentum space technique because it requires the multiloop
integration of entangled momenta. Within the configuration space technique the
generalization to any number of lines (or loops) is immediate by no effort.
Consider first a three-loop case of water melon diagrams (also called banana
diagrams~\cite{Broadhurst2} or basketball diagrams~\cite{Braaten}). The
polarization function of the equal mass water melon diagram with four
propagators in $D=4$ space-time is given by
\begin{equation}
\Pi(x)=\frac{m^4 K_1(mx)^4}{(2\pi)^8x^4}.
\end{equation}
The exact spectral density of this diagram can be obtained from
Eq.~(\ref{exactrho4}) while the near threshold expansion can be found using 
Eq.~(\ref{serint}). We construct the expansion of the spectral density near
threshold explicitly and compare it with the exact result. The expansion of
the integrand (cf.\ Eq.~(\ref{pidas})) reads
\begin{eqnarray}
\lefteqn{\pi^2\pfrac{ipx}2^{-1}H_{1,N}^{as}(px)\Pi_N^{as}(x)x^{3+2\eps}
  \ =\ \frac{m^2e^{(p-4m)x}}{(4\pi)^4\sqrt{2\pi}p^{3/2}}x^{-9/2+2\eps}\
  \times}\nonumber\\&&\qquad\times\
  \left\{1+\frac3{2mx}-\frac3{8px}+\frac3{8m^2x^2}-\frac{15}{128p^2x^2}
  -\frac9{16mpx^2}+O(x^{-3})\right\}.
\end{eqnarray}
After the integration and the calculation of the discontinuity we obtain the
expansion of the spectral density in the form
\begin{eqnarray}\label{pidas440}
\lefteqn{\tilde\rho(E)\ =\ \frac{E^{7/2}M^{1/2}}{26880\pi^5\sqrt2}
\Bigg\{1-\frac14\eta+\frac{81}{352}\eta^2-\frac{2811}{18304}\eta^3
  +\frac{17581}{292864}\eta^4}\\&&\kern-16pt
  +\frac{1085791}{19914752}\eta^5-\frac{597243189}{3027042304}\eta^6
  +\frac{4581732455}{12108169216}\eta^7
  -\frac{496039631453}{810146594816}\eta^8+O(\eta^9)\Bigg\}\nonumber
\end{eqnarray}
where $\eta=E/M$ and $M=4m$ is the threshold value. One sees the difference
with the previous three-line case. In Eq.~(\ref{pidas440}) the cut represents 
the square root branch while in the three-line case it was a logarithmic cut. 
One can easily figure out the reason for this by looking at the asymptotic
structure of the integrand. For even number of lines (i.e.\ odd number of
loops) it is a square root branch, while for an odd number of lines (even
number of loops) it is a logarithmic branch. This is true in even space-time
dimensions. In the general case the structure of the cut depends on the
dimensionality of the space-time as well. The general formula reads
\begin{equation}\label{genthr}
\tilde\rho(E)\sim E^{(\lambda+1/2)(n-1)-1}(1+O(E)).
\end{equation}
For $D=4$ space-time dimension (i.e.\ $\lambda=1$) we can verify the result of
Ref.~\cite{GKP1} (cf.\ Eq.~(\ref{genthr})),
\begin{equation}
\tilde\rho(E)\sim E^{(3n-5)/2}(1+O(E)).
\end{equation}

Numerically Eq.~(\ref{pidas440}) reads
\begin{eqnarray}\label{sernum}
\lefteqn{\tilde\rho(E)\ =\ 8.5962\cdot 10^{-5}E^{7/2}M^{1/2}
   \Big\{1.000-0.250\eta+0.230\eta^2}\\&&
  -0.154\eta^3+0.060\eta^4+0.055\eta^5-0.197\eta^6+0.378\eta^7-0.612\eta^8
  +O(\eta^9)\Big\}\nonumber
\end{eqnarray}
where we have written down the coefficients up to three decimal places. It is
difficult to say anything definite about the convergence of this series. By
construction it is an asymptotic series. However, we stress that the practical
(or explicit) convergence can always be checked by comparing series expansions
like the one shown in Eq.~(\ref{sernum}) with the exact spectral density given
in Eq.~(\ref{exactrho4}) by numerical integration. 

We conclude this part of the paper by the statement that the spectral density
of the water melon diagram is most efficiently calculable within the
configuration space technique. Whether it is the exact result or the
expansion, the configuration space technique can readily deliver the desired
result. The exact formula in Eq.~(\ref{exactrho4}) as well as the threshold
expansion obtained from it can be used to calculate the spectral density for
an arbitrarily large number of internal lines. We stress that the case of
different masses does not lead to any complications within the configuration
space technique: the exact formula in Eq.~(\ref{exactrho}) and/or the near
threshold expansion work equally well for any arrangement of masses. We do not
present plots for general cases of different masses because they are not very
illustrative, showing only the common threshold. However, there is some
interesting kinematic regime for different masses which is important for
applications and which, to our best knowledge, have not been touched earlier.
An analytical solution for the expansion of the spectral density in this
regime is given in the next section.

\begin{figure}[ht]
\epsfig{figure=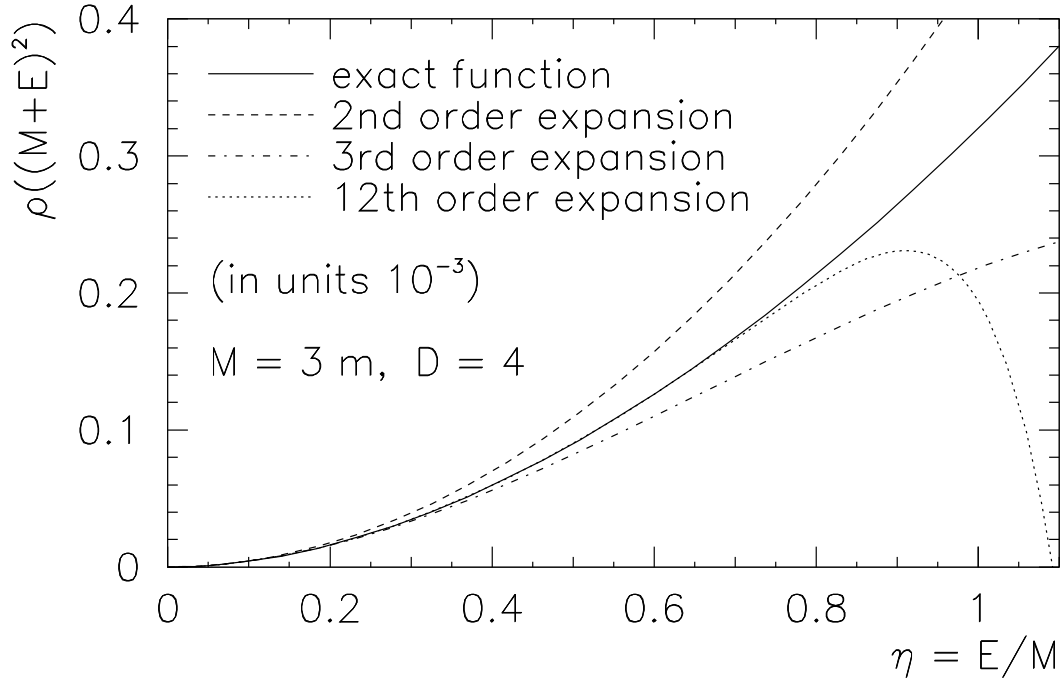, height=9truecm, width=14truecm}
\caption{\label{fig2}Various results for the spectral density for $n=3$ equal
masses in $D=4$ space-time dimensions in dependence on the threshold parameter
$E/M$. Shown are the exact solution obtained by using Eq.~(\ref{exactrho4})
(solid curve) and threshold expansions for different orders taken from
Eq.~(\ref{pidas430}) (dashed to dotted curves).}
\end{figure}

\section{Strongly asymmetric case $m_0\ll M$}
The threshold expansion for equal (or close) masses breaks down for
$E\approx M=\sum m_i$. The example is shown in Fig.~\ref{fig2} for the $D=4$
proper sunset. However, if the masses are not equal, the region of the
break-down of the expansion is determined by the mass with the smallest
numerical value. The simplest example where one can see this phenomenon is the
analytical expression for the spectral density of the simple loop (degenerate
water melon diagram) with two different masses $m_1$ and $m_2$. In $D=4$
space-time dimensions (see e.g.\ Ref.~\cite{GKP1}) one has 
\begin{equation}\label{rho42m}
\tilde\rho(E)=\frac{\sqrt{E(E+2m_1)(E+2m_2)(E+2M)}}{(4\pi(M+E))^2}
\end{equation}
where $M=m_1+m_2$. The threshold expansion is obtained by expanding the right
hand side of Eq.~(\ref{rho42m}) in $E$ for small values of $E$. If $m_2$ is
much smaller than $m_1$, the expansion breaks down at $E\approx 2m_2$. The
break-down of the series expansion can also be observed in more general cases.
If one of the masses (which we call $m_0$) is much smaller than the other
masses, the threshold expansion is only valid in a very limited region
$E\simle 2m_0$.

To generalize the expansion and extend it to the region of $E\sim M$ one has
to treat the smallest mass exactly. In this case one can use a method which
we call the resummation of the smallest mass contributions. Below we describe
the resummation technique. We start with the representation
\begin{equation}\label{pipas}
\Pi^{pas}(p)=\pi^{\lambda+1}\int\pfrac{ipx}2^{-\lambda}
  H_{\lambda,N}^{-as}(ipx)\Pi^{as}_{m_0}(x)x^{2\lambda+1+2\eps}dx
\end{equation}
which is the part of the polarization function contributing to the spectral
density. The integrand in Eq.~(\ref{pipas}) has the form
\begin{equation}
\Pi^{as}_{m_0}(x)=\Pi^{as}_{n-1}(x)D(m_0,x)
\end{equation}
where the asymptotic expansions are substituted for all the propagators except
for the one with the small mass $m_0$. This is indicated by the index
``{\it pas}'' in Eq.~(\ref{pipas}) which stands for ``partial asymptotic''.
The main technical observation leading to the generalization of the expansion
method is that $\Pi^{pas}(p)$ is still analytically computable in a closed
form. Indeed, the genuine integral to compute has the form 
\begin{eqnarray}\label{intm0}
\lefteqn{\int_0^\infty x^{\mu-1}e^{-\tilde\alpha x}K_\nu(\beta x)dx
  \ =}\nonumber\\
  &=&\frac{\sqrt\pi(2\beta)^\nu}{(2\tilde\alpha)^{\mu+\nu}}
  \frac{\Gamma(\mu+\nu)\Gamma(\mu-\nu)}{\Gamma(\mu+1/2)}\
  {_2F_1}\left(\frac{\mu+\nu}2,\frac{\mu+\nu+1}2;\mu+\frac12;
  1-\frac{\beta^2}{\tilde\alpha^2}\right)
\end{eqnarray}
where $\tilde\alpha=\Delta-m_0$ and $\beta=m_0$. The integral $\Pi^{pas}(p)$
in Eq.~(\ref{pipas}) is thus expressible in terms of hypergeometric
functions~\cite{GradshteynRyzhik,AbramowitzStegun}. For constructing the
spectral density, being our main concern as mentioned before, one has to find
the discontinuity of the right hand side of Eq.~(\ref{intm0}). There are
several ways to do this. We proceed by applying the discontinuity operation to
the integrand of the integral representation of the hypergeometric function.
The resulting integrals are calculated again in terms of hypergeometric
functions. Indeed,
\begin{eqnarray}\label{genuine}
\lefteqn{\frac1{2\pi i}\Disc\int_0^\infty x^{\mu-1}e^{\alpha x}
  K_\nu(\beta x)dx\ =}\nonumber\\
  &=&\frac{2^\mu(\alpha^2-\beta^2)^{1/2-\mu}}{\alpha^{1/2-\nu}\beta^\nu}
  \frac{\Gamma(3/2)}{\Gamma(3/2-\mu)}\ {_2F_1}\left(\frac{1-\mu-\nu}2,
  \frac{2-\mu-\nu}2;\frac32-\mu;1-\frac{\beta^2}{\alpha^2}\right)
\end{eqnarray}
where $\alpha=E+m_0$. The final expression in Eq.~(\ref{genuine}) completely
solves the problem of the generalization of the near threshold expansion
technique. For integer values of $\mu$ there are no singular Gamma functions
(with negative integer argument). Therefore we can lift up the regularization 
and set $\eps=0$ when using this expression. We thus have found a direct
transition from the polarization function as expressed through the integral to
the spectral density in terms of one hypergeometric function for each genuine
integral. There is no need to use the recurrence relations available for
hypergeometric functions.

In the following subsections we give explicit examples for $D=4$ and compare
with the exact result in Eq.~(\ref{exactrho4}) and the pure expansion near the 
threshold. In the following the standard threshold expansion without
resummation is called the pure threshold expansion.

\begin{figure}[ht]
\epsfig{figure=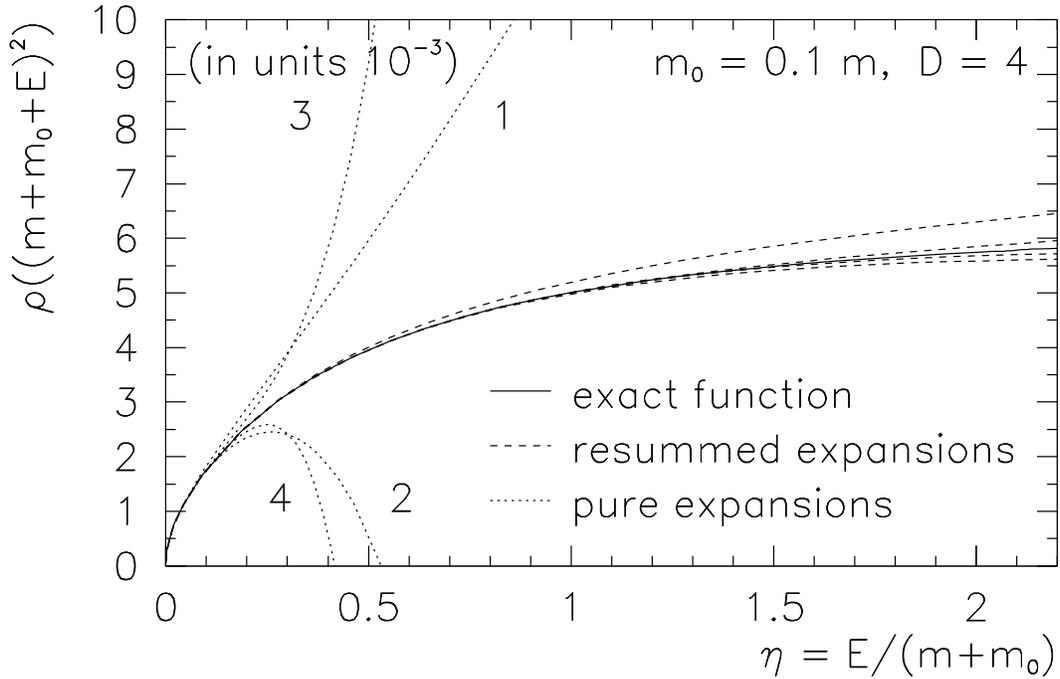, height=9truecm, width=14truecm}
\caption{\label{fig3}Various solutions for the spectral density for two
masses $m$ and $m_0\ll m$ and $D=4$ space-time dimensions. Shown are the exact
solution which is obtained by using Eq.~(\ref{exactrho4}) (solid curve), the
pure threshold expansions using Eq.~(\ref{pidas42t}) (dotted curves), and the
solutions for the resummation of the smallest mass contributions like in
Eq.~(\ref{pipas42t}) (dashed curves), both expansions from the first up to the
fourth order in the asymptotic expansion. For the pure threshold expansion the
order is indicated explicitly.}
\end{figure}

\subsection{The two-line water melon with a small mass}
We start with a (over)simplified example of the two-line diagram with masses
$m$ and $m_0\ll m$ in four space-time dimensions. In this example the
expansion of the spectral density and its generalized expansion can be readily
compared analytically with the exact result in Eq.~(\ref{rho42m}). This is
the feature that justifies our discussion in this section. The results for the
spectral density of this diagram are shown in Fig.~\ref{fig3}. The solid curve
displays the exact result obtained by using Eq.~(\ref{exactrho4}) (which
reproduces the analytical expression in Eq.~(\ref{rho42m})). We compare this
result with the two expansions. 

The pure expansion of the spectral density near threshold (the second order
asymptotic expansion should suffice to show the general features in a short
and concise form) is given by
\begin{eqnarray}\label{pidas42t}
\tilde\rho^{das}(E)&=&\frac{\sqrt{2m_0mE}}{8\pi^2M^{3/2}}
  \Bigg\{1+\left(\frac1m+\frac1{m_0}-\frac7M\right)\frac{E}4\nonumber\\&&
  -\left(\frac1{m_0^2}+\frac1{m^2}+\frac{12}{m_0m}-\frac{79}{M^2}\right)
  \frac{E^2}{32}+O(E^3)\Bigg\}
\end{eqnarray}
where $M=m+m_0$. As mentioned above, this series breaks down for $E>2m_0$
(see Eq.~(\ref{rho42m})). If we look at the dotted curves in Fig.~\ref{fig3}
this becomes obvious. Here we have plotted the series expansions up to the
fourth order with the mass arrangement $m_0=m/10$. The dashed lines represent
the resummation of the smallest mass contributions. The analytical expression
for the spectral density of the polarization function in Eq.~(\ref{pipas})
for the generalized asymptotic expansion based on Eq.~(\ref{genuine})
is given by
\begin{eqnarray}\label{pipas42t}
\lefteqn{\tilde\rho^{pas}(E)\ =\ \frac{\sqrt{mE(E+2m_0)}}{8\pi^2(E+M)^{3/2}}
  \Bigg\{{_2F_1}\left(0,\frac12;\frac32;1-\frac{m_0^2}{(E+m_0)^2}\right)}
  \nonumber\\&&+\frac{E(E+2m_0)}{8m(E+M)}\
  {_2F_1}\left(\frac12,1;\frac52;1-\frac{m_0^2}{(E+m_0)^2}\right)\\&&
  -\frac{E^2(E+2m_0)^2}{128m^2(E+M)^2}\left(1+\frac{16m(E+M)}{5(E+m_0)^2}
  \right)\
  {_2F_1}\left(1,\frac32;\frac72;1-\frac{m_0^2}{(E+m_0)^2}\right)+\ldots
  \Bigg\}.\nonumber
\end{eqnarray}
We have set the regularization parameter $\eps=0$ because the spectral density
is finite. With $\eps=0$ the resulting expressions for the hypergeometric 
functions in Eq.~(\ref{genuine}) simplify. The first term in the curly
brackets of Eq.~(\ref{pipas42t}) is obviously equal to $1$ in this limit
because the first parameter of the hypergeometric function vanishes for
$\eps=0$. However, we keep Eq.~(\ref{pipas42t}) in its given form to show the
structure of the contributions. The generalized threshold expansion has the
form
\begin{eqnarray}\label{genStr}
\tilde\rho^{pas}(E)=g_0(E,m_0)+Eg_1(E,m_0)+E^2g_2(E,m_0)+\ldots
\end{eqnarray}
where the functions $g_j(E,m_0)$ represent effects of the resummation of the
smallest mass and are not polynomials in the threshold parameter parameter
$E$. In the simple two-line case the hypergeometric functions reduce to
elementary functions. For instance,
\begin{eqnarray}
\label{resumSm}
\lefteqn{{_2F_1}\left(\frac12,1;\frac52;1-\frac{m_0^2}{(E+m_0)^2}\right)
  \ =}\\
  &=&\frac{3(E+m_0)}{2E(E+2m_0)}\left(E+m_0-\frac{m_0^2}{2\sqrt{E(E+2m_0)}}
  \ln\pfrac{E+m_0+\sqrt{E(E+2m_0)}}{E+m_0-\sqrt{E(E+2m_0)}}\right).\nonumber
\end{eqnarray}
Higher order contributions are given by hypergeometric functions with larger
numerical values of the parameters. They can be simplified by using Gaussian
recurrence relations for hypergeometric functions (see e.g.\
Ref.~\cite{GradshteynRyzhik}).

\begin{figure}[ht]
\epsfig{figure=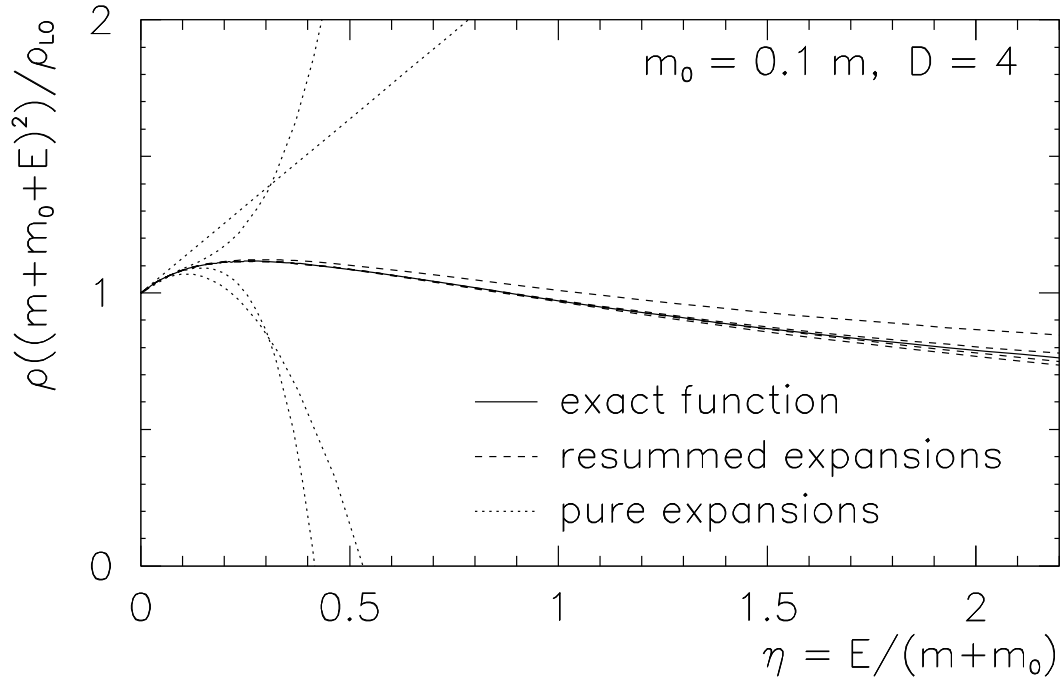, height=9truecm, width=14truecm}
\caption{\label{fig4}The same as Fig.~\ref{fig3} where 
the spectral density
is normalized to the leading order expression of the pure threshold
expansion.}
\end{figure}

The convergence of the expansion in Eq.~(\ref{pipas42t}) breaks down only at
$E\sim M=m+m_0$. The resummation leads to an essential improvement of the
convergence in comparison with the pure threshold expansion. In
Fig.~\ref{fig4} we show the same curves divided by the leading order term.
This representation is more convenient for the diagrams which we will discuss
in following subsections.

Note that Eq.~(\ref{resumSm}) does not lead to the exact function in
Eq.~(\ref{rho42m}) because terms of order $E^N$ stemming from the difference
part $\Pi^{di}(p)$ of the correlator are missing. It simply corrects the
behaviour of the coefficient functions by the small mass contributions.

\begin{figure}[ht]
\epsfig{figure=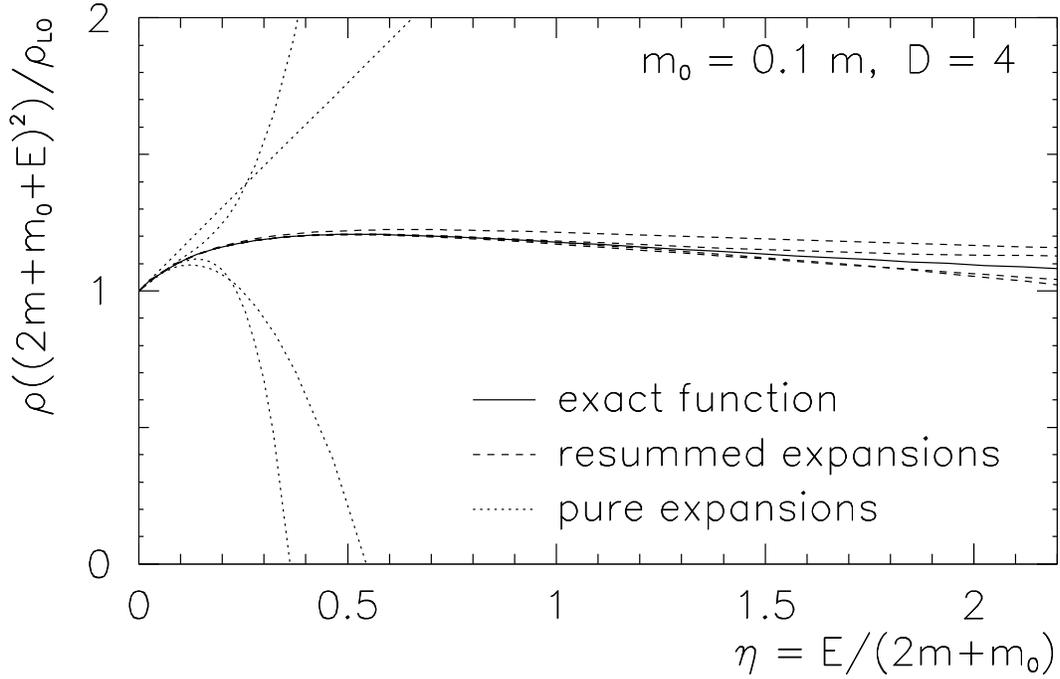, height=9truecm, width=14truecm}
\caption{\label{fig5}The spectral density for the sunset diagram in $D=4$
space-time dimensions with a tiny mass $m_0$, normalized to the general power
behaviour. Shown are the exact result obtained by using Eq.~(\ref{exactrho4})
(solid curve), the threshold expansion according to Eq.~(\ref{pidas43t})
(dotted curves), and the result for the resummation of the smallest mass
contributions according to Eq.~(\ref{pipas43t}) (dashed curves).}
\end{figure}

\subsection{The sunset diagram with a small mass}
Here we analyze the sunset diagram with two equal masses $m$ and a third mass
$m_0\ll m$ ($m_0=m/10$). The exact result obtained by using
Eq.~(\ref{exactrho4}) and normalized to the leading order term is shown in
Fig.~\ref{fig5} as the solid curve. The pure expansion near the threshold
reads
\begin{eqnarray}\label{pidas43t}
\tilde\rho^{das}(E)&=&\frac{mE^2\sqrt{m_0M}}{128\pi^3M^2}
  \Bigg\{1+\left(\frac1{m_0}+\frac2m-\frac{13}M\right)\frac{E}8\nonumber\\&&
  -\left(\frac5{m_0^2}+\frac4{m^2}+\frac{39}{m_0m}+\frac{153}{mM}
  -\frac{1115}{M^2}\right)\frac{E^2}{512}+O(E^3)\Bigg\}.
\end{eqnarray}
It is shown by the dotted curves in Fig.~\ref{fig5}. In case of the
resummation of the smallest mass contributions we obtain hypergeometric
functions which do not obviously reduce to elementary functions in this case. 
The result for the spectral density within the asymptotic expansion up to the
second order in Eq.~(\ref{pipas}) is given by
\begin{eqnarray}\label{pipas43t}
\lefteqn{\tilde\rho^{pas}(E)
  =\frac{mE^2(E+2m_0)^2}{512\pi^3(E+m_0)^{3/2}(E+M)^{3/2}}
  \Bigg\{{_2F_1}\left(\frac34,\frac54;3;1-\frac{m_0^2}{(E+m_0)^2}\right)}
  \nonumber\\&&
  +\frac{E(E+2m_0)}{8m(E+M)}\left(1+\frac{3m}{2(E+m_0)}\right)\
  {_2F_1}\left(\frac54,\frac74;4;1-\frac{m_0^2}{(E+m_0)^2}\right)\nonumber\\&&
  -\frac{E^2(E+2m_0)^2}{512m^2(E+M)^2}
  \left(1+\frac{5m}{2(E+m_0)}\right)
  \left(1+\frac{9m}{2(E+m_0)}\right)\times\nonumber\\&&\qquad
  {_2F_1}\left(\frac74,\frac94;5;1-\frac{m_0^2}{(E+m_0)^2}\right)\Bigg\}.
\end{eqnarray}
We see that the dashed curves in Fig.~\ref{fig5} that represent the result in
Eq.~(\ref{pipas43t}) approximate the exact curve much better than the dotted
curves.

\begin{figure}[ht]
\epsfig{figure=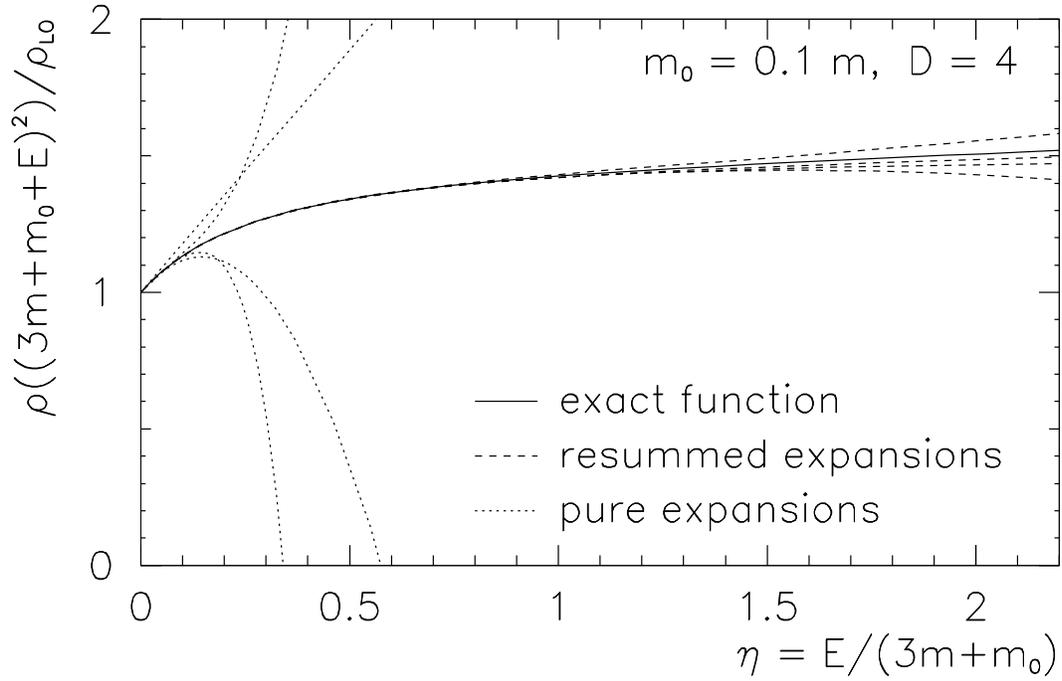, height=9truecm, width=14truecm}
\caption{\label{fig6}The spectral density for the four-line water melon
diagram in $D=4$ space-time dimensions with a tiny mass $m_0$, normalized to
the general power behaviour. Shown are the exact result obtained by using
Eq.~(\ref{exactrho4}) (solid curve), the threshold expansion according to
Eq.~(\ref{pidas44t}) (dotted curves), and the result for the resummation of
the smallest mass contributions according to Eq.~(\ref{pipas44t}) (dashed
curves).}
\end{figure}

\subsection{The four-line water melon with a small mass}
With this example we conclude our consideration of the strongly asymmetric
case and at the same time show the way to the multi-line water melon diagrams
which can be treated in an analogous manner. The result for the exact
expression obtained by using Eq.~(\ref{exactrho4}) is shown in Fig.~\ref{fig6}
as a solid line, normalized to the leading order term. The dotted lines
represent the results for the pure expansion near threshold which is given by
\begin{eqnarray}\label{pidas44t}
\tilde\rho^{das}(E)&=&\frac{m^{3/2}E^{7/2}\sqrt{2m_0}}{3360\pi^5 M^{3/2}}
  \Bigg\{1+\left(\frac1{m_0}+\frac3m-\frac{19}M\right)\frac{E}{12}\\&&\qquad
  -\left(\frac5{m_0^2}-\frac3{m^2}+\frac{28}{m_0m}+\frac{368}{mM}
  -\frac{2195}{M^2}\right)\frac{E^2}{1056}+O(E^3)\Bigg\}.\nonumber
\end{eqnarray}
The asymptotic expansion to the second order in Eq.~(\ref{pipas}) gives
\begin{eqnarray}\label{pipas44t}
\lefteqn{\tilde\rho^{pas}(E)
  \ =\ \frac{m^{3/2}E^{7/2}(E+2m_0)^{7/2}}{26880\pi^5(E+m_0)^3(E+M)^{3/2}}
  \Bigg\{{_2F_1}\left(\frac32,2;\frac92;1-\frac{m_0^2}{(E+m_0)^2}\right)}
  \nonumber\\&&
  +\frac{E(E+2m_0)}{8m(E+M)}\left(1+\frac{8m}{3(E+m_0)}\right)\
  {_2F_1}\left(2,\frac52;\frac{11}2;1-\frac{m_0^2}{(E+m_0)^2}\right)
  \nonumber\\&&
  +\frac{E^2(E+2m_0)^2}{1408(E+M)^2}\left(1-\frac{32m^2}{3(E+m_0)^2}\right)\
  {_2F_1}\left(\frac52,3;\frac{13}2;1-\frac{m_0^2}{(E+m_0)^2}\right)\Bigg\}.
  \qquad\qquad
\end{eqnarray}
In Fig.~\ref{fig6} one can see how the expansion improves if the resummation
of the smallest mass contributions (displayed as dashed lines) is performed.

The result of this section is quite general and applicable to all cases of one
small mass. For some particular arrangement of masses one can obtain even
simpler expressions as discussed in the next section.

\subsection{The convolution with a small mass}
In this section we obtain a result for the resummation of the smallest mass
effects along a different route, namely, via the convolution of spectral
densities. However, this method works in a narrower kinematic region than the
method described in the previous section. In $D=4$ space-time dimensions, the
convolution weight is given by
\begin{equation}
\rho(s;s_1;s_2)=\frac1{(4\pi)^2s}\sqrt{(s-s_1-s_2)^2-4s_1s_2}.
\end{equation}
The upper limit of the integration is determined by the requirement of
positivity of the the square root argument. The zeros of the square root with
respect to $s_2$ are given by $s_2^\pm=(\sqrt s\pm\sqrt s_1)^2$, and the
demand $(s_2-s_2^+)(s_2-s_2^-)>0$ together with $s_2^+>s_2^-$ leads to
$s_2>s_2^+$ or $s_2<s_2^-$. The physical region is the latter one. With
$\rho_1(s)=\delta(s-m_0^2)$ for the spectral density of the single small mass
line we obtain
\begin{eqnarray}
\rho(s)&=&\int_0^\infty ds_1\int_{M'^2}^{(\sqrt s-\sqrt{s_1})^2}ds_2
  \rho(s;s_1;s_2)\rho_1(s_1)\rho_2(s_2)\ =\nonumber\\
  &=&\frac1{(4\pi)^2s}\int_{M'^2}^{(\sqrt s-m_0)^2}
  \sqrt{(s-m_0^2-s_2)^2-4m_0^2s_2}\ \rho_2(s_2)ds_2
\end{eqnarray}
where the low limit of integration is $M'=M-m_0$. We insert $s=(M+E)^2$ and
$s_2=(M'+E')^2$ and obtain
\begin{eqnarray}\label{rhoconv}
\tilde\rho(E)&=&\frac1{(4\pi)^2(M+E)^2}\int_0^E\sqrt{(E-E')(E+E'+2M)+m_0^2}
  \times\nonumber\\&&\qquad\times\
  \sqrt{(E-E'+2m_0)(E+E'+2M')+m_0^2}\ \frac{\tilde\rho'(E')dE'}{2(M'+E')}
\end{eqnarray}
where $\tilde\rho'(E')=\rho_2((M'+E')^2)$. For this function we use the
threshold expansion in $E'/M'$ as expansion parameter. For small $E<M'$ 
the threshold expansion inserted for $\tilde\rho'(E')$ is valid because 
$E<M'$ implies $E'<M'$. The described procedure can be extended to the case of
a very light sub-block of the diagram, e.g.\ a light fish diagram. In this
case we have to replace $\rho_1(s)$ by the spectral density of the light
sub-diagram which is well-known.

\section{Recovering $\Pi(p)$ through $\rho(s)$ near threshold}
The analytic structure of water melon diagrams is completely fixed by the
dispersion representation. Therefore we have focussed on the computation of
the spectral density as the basic quantity important both for applications and
the theoretical investigation of the diagram. However, with an analytical
expression for the spectral density $\rho(s)$ at hand we can readily
reconstruct the non-analytic piece of the polarization function in momentum
space by using the dispersion relation
\begin{equation}
\Pi(p)=\int\frac{\rho(s)ds}{s-p^2}.
\end{equation}
We rewrite this equation in terms of threshold parameters according to
$p=M-\Delta$ and $s=(M+E)^2$ and obtain
\begin{equation}
\tilde\Pi(\Delta)\equiv\Pi(M-\Delta)=\int_0^\infty
  \frac{2(M+E)\tilde\rho(E)dE}{(E+\Delta)(2M+E-\Delta)}.
\end{equation}
UV singularities can be removed by subtraction or by dimensional
regularization. We again use the unorthodox dimensional regularization
prescription. For a general form of the threshold expansion
$\tilde\rho(E)=E^\gamma\sum a_k E^k$ we have to calculate integrals of the
form
\begin{eqnarray}\label{restore}
\tilde\Pi^\sigma(\Delta)&=&\int_0^\infty\frac{E^\sigma dE}{(E+\Delta)
  (2M+E-\Delta)}=-\frac\pi{\sin(\pi\sigma)}
  \frac{\Delta^\sigma-(2M-\Delta)^\sigma}{2(M-\Delta)}.
\end{eqnarray}
Only the powers $\Delta^\sigma$ contribute to the singular part of the
polarization function. Expressions like the one presented in
Eq.~(\ref{restore}) then allow one to restore that part of the polarization
function $\Pi(p)$ which has singularities near the threshold.

\begin{figure}[ht]
\epsfig{figure=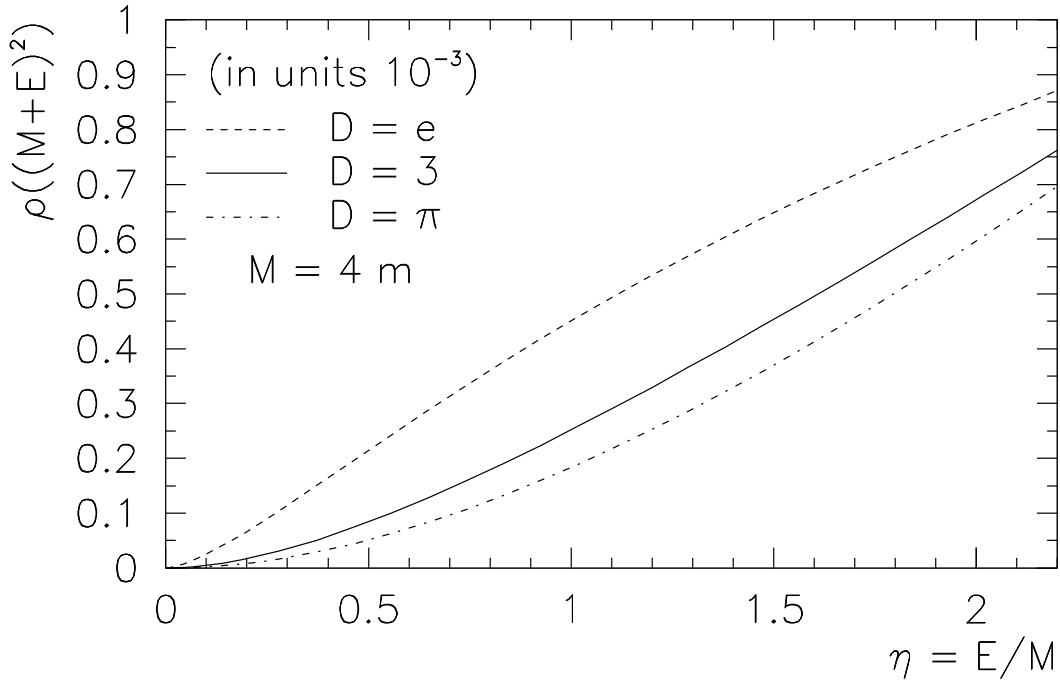, height=9truecm, width=14truecm}
\caption{\label{fig7}The spectral density for the four-line water melon
diagram with equal masses for $D=e=2.718\ldots$, $D=3$, and $D=\pi=3.14\ldots$
space-time dimensions.}
\end{figure}

\section{Conclusion}
We have discussed the configuration space technique for the calculation of
$n$-line two-point diagrams termed water melon diagrams. This technique allows
one to calculate the spectral density for arbitrary space-time dimensions and
any number of internal lines with arbitrary mass values. Within this technique
one can use either an exact representation as one-fold integral or an
expansion near the production threshold. We have developed an efficient method
for constructing the near threshold expansion of water melon diagrams that
uses only asymptotic expansions of Bessel functions which are well-known and
simple functions. We have considered a strongly asymmetric mass arrangement 
where one mass is much smaller than the others. In this case the ``pure''
threshold expansion which is done in terms of the threshold parameter $E$
breaks down at the energies in the vicinity of the smallest mass. In order to
extend the approximation to higher energies we have introduced a generalized
threshold expansion which exactly resums all contributions of the smallest
mass. We have presented closed expressions for the generalized expansion in
several particular cases and demonstrated the improvement of the convergence
gained by the resummation of the smallest mass contributions. The particular
kinematic regime of this case could be treated analytically because it reduced
to the evaluation of the one-fold integral in terms of hypergeometric
functions. The discontinuity across the physical cut for the generalized
expansion has been found in terms of hypergeometric functions as well.

To conclude, we stress that the configuration space technique is a powerful
and convenient tool for investigating different properties of water melon
diagrams. The practical convenience of our method is demonstrated in 
Fig.~\ref{fig7} where we have plotted the spectral density for a four-line
water melon diagram in $D=e=2.718\ldots$, $D=3$, and $D=\pi=3.14\ldots$
space-time dimensions.

\vspace{3mm}\noindent
{\large\bf Acknowledgments}\\[2mm]
We thank Andrei Davydychev and J\"urgen K\"orner for interesting discussions
and careful reading the manuscript. A.A.~Pivovarov acknowledges a valuable
discussion with V.A.~Smirnov on the region separation technique in threshold
expansions. The work of A.A.~Pivovarov is supported in part by the Volkswagen
Foundation under contract No.~I/73611 and by the Russian Fund for Basic
Research under contract 99-01-00091. S.~Groote gratefully acknowledges a grant
given by the Graduiertenkolleg ``Eichtheorien -- experimentelle Tests und
theoretische Grundlagen''.


\begin{thebibliography}{99}
\bibitem{PDG}Particle Data Group, ``Review of Particle Properties'',
  Eur.~Phys.~J. {\bf C3} (1998) 1
\bibitem{revloop}K.G.~Chetyrkin, J.H.~K\"uhn and A.~Kwiatkowski,
  Phys.~Rep.\ {\bf 277} (1996) 189  
\bibitem{ChKSt2}K.G.~Chetyrkin, J.H.~K\"uhn and M.~Steinhauser,
  Nucl.~Phys.\ {\bf B505} (1997) 213
\bibitem{Broadhurst1}D.J.~Broadhurst, Eur.~Phys.~J.\ {\bf C8} (1999) 311
\bibitem{Osland}O.M.~Ogreid, P.~Osland, ``More series related to the Euler
  series'',\\ Report No.~BERGEN-1999-04, hep-th/9904206 
\bibitem{renZ}R.~Coquereaux, Phys.~Rev.\ {\bf D23} (1981) 2276;\\
  V.I.~Zakharov, Nucl.~Phys.\ {\bf B385} (1992) 452;\\
  D.J.~Broadhurst and A.G.~Grozin, Phys.~Rev.\ {\bf D52} (1995) 4082
\bibitem{GKP1}S.~Groote, J.G.~K\"orner and A.A.~Pivovarov,
  Nucl.~Phys.\ {\bf B542} (1999) 515; Eur.~Phys.~J.\ {\bf C11} (1999) 279;
  Phys.~Lett.\ {\bf 443 B} (1998) 269
\bibitem{Berends}F.A.~Berends, M.~Buza, M.~B\"ohm and R.~Scharf, 
  Z.~Phys.\ {\bf C63} (1994) 227
\bibitem{Ussyukina}F.A.~Berends, A.I.~Davydychev and N.I.~Ussyukina,
  Phys.~Lett.\ {\bf 426 B} (1998) 95
\bibitem{Jackiw}S.~Coleman and E.~Weinberg, Phys.~Rev.\ {\bf D7} (1973) 1888;\\
  R.~Jackiw, Phys.~Rev.\ {\bf D9} (1974) 1686;\\
  R.~Jackiw and S.Templeton, Phys.~Rev.\ {\bf D23} (1981) 2291
\bibitem{Rajantie}A.K.~Rajantie, Nucl.~Phys.\ {\bf B480} (1996) 729
\bibitem{Chung}J.M. Chung and B.K. Chung, ``Renormalization group improvement
  of the effective potential in massive $\Phi^4$ theory: NNNLO logarithm
  resummation'', Report No.~MIT-CTP-2929, hep-th/9911196
\bibitem{Braaten}J.O.~Andersen, E.~Braaten and M.~Strickland,
  ``The massive thermal basketball diagram'', hep-ph/0002048 
\bibitem{Chung2}J.M.~Chung and B.K.~Chung,
  Phys.~Rev.\ {\bf D59} (1999) 105014
\bibitem{temp}D.J.~Gross, R.D.~Pisarski and L.G.~Yaffe,
  Rev.~Mod.~Phys.\ {\bf 53} (1981) 43;\\
  T.~Appelquist and R.D.~Pisarski, Phys.~Rev.\ {\bf D23} (1981) 2305
\bibitem{Ioffe}B.L.~Ioffe, Nucl.~Phys.\ {\bf B188} (1981) 317 
\bibitem{Kras}N.V.~Krasnikov, A.A.~Pivovarov and A.N.~Tavkhelidze,\\
  Z.~Phys.\  {\bf C19} (1983) 301; JETP Lett.\  {\bf 36} (1982) 333
\bibitem{bar}A.A.~Ovchinnikov, A.A.~Pivovarov and L.R.~Surguladze,\\
  Yad.~Fiz.\ {\bf 48} (1988) 562; Int.~J.~Mod.~Phys.\ {\bf A6} (1991) 2025
\bibitem{barGro}S.~Groote, J.~G.~K\"orner and A.~A.~Pivovarov,
  Phys.~Rev.\ {\bf D61} (2000) 071501
\bibitem{NcHoo}G.~'t Hooft, Nucl.~Phys.\ {\bf B72} (1974) 461 
\bibitem{NcBar}E.~Witten, Nucl.~Phys.\ {\bf B160} (1979) 57
\bibitem{dibar}S.A.~Larin {\em et al.}, Yad.~Fiz.\ {\bf 44} (1986) 1066
\bibitem{nucl}S.A.~Kulagin and A.A.~Pivovarov, Yad.~Fiz.\ {\bf 45} (1987) 952
\bibitem{Post}P.~Post and K.~Schilcher, Phys.~Rev.~Lett.\ {\bf 79} (1997) 4088
\bibitem{Gasser}J.~Gasser and M.E.~Sainio, Eur.~Phys.~J.\ {\bf C6} (1999) 297 
\bibitem{Narison}S.~Narison and A.A.~Pivovarov, 
  Phys.~Lett.\ {\bf 327 B} (1994) 341
\bibitem{hyb1}I.I.~Balitsky, D.I.~D'Yakonov and A.V.~Yung,
  Phys.~Lett.\ {\bf 112 B} (1982) 71
\bibitem{Broadhurst2}D.J.~Broadhurst, Z.~Phys.\ {\bf C54} (1992) 599
\bibitem{Baikov}P.A.~Baikov, ``Advanced method of solving recurrence
  relations for multiloop Feynman integrals'', talk presented at the
  $6^{\rm th}$ International Workshop on New Computing Techniques in Physics
  Research (AIHENP 99), Heraklion, Greece, 12--16 April, 1999
\bibitem{Ligterink}N.E.~Ligterink, ``Solving multiloop Feynman diagrams using
  light front coordinates'', Report No.~ECT-99-016, hep-ph/9911411
\bibitem{Smirnov}A.I.~Davydychev and V.A.~Smirnov,
  Nucl.~Phys.\ {\bf B554} (1999) 391
\bibitem{KasteningKleinert}B.~Kastening, H.~Kleinert, ``Efficient algorithm
  for perturbative calculation of multiloop Feynman integrals'',
  quant-ph/9909017
\bibitem{Watson}G.N.~Watson, ``Theory of Bessel functions'', Cambridge, 1944
\bibitem{Mendels}E.~Mendels, Nuovo~Cim.\ {\bf 45 A} (1978) 87
\bibitem{Meijer}C.S.~Meijer, Proc.~Amsterdam~Akad.~Wet.\ (1940) 599; 702
\bibitem{Erdelyi}A.~Erdelyi (Ed.), ``Tables of integral transformations'',\\
  Volume~2, Bateman manuscript project, 1954
\bibitem{Penin}A.H.~Hoang {\em et al.}, ``Top-antitop pair production close
  to threshold: Synopsis of recent NNLO  results,''
  Report No.~SLAC-PUB-8369, hep-ph/0001286;\\
  A.~Czarnecky and K.~Melnikov, Phys.~Rev.~Lett.\ {\bf 80} (1998) 2531;\\
  M.~Beneke, A.~Signer and V.A.~Smirnov,
  Phys.~Rev.~Lett.\ {\bf 80} (1998) 2535;\\
  K.~Melnikov and A.~Yelkhovsky, Nucl.~Phys.\ {\bf B528} (1998) 59;\\
  A.A.~Penin and A.A.~Pivovarov, hep-ph/9904278;
  Nucl.~Phys.\ {\bf B549} (1999) 217;  Nucl.~Phys.\ {\bf B550} (1999) 375;
  Phys.~Lett.\ {\bf 443 B} (1998) 264;\\
  O.I.~Yakovlev, Phys.~Lett.\ {\bf 457 B} (1999) 170;\\
  A.H.~Hoang and T.~Teubner, Phys.~Rev.\ {\bf D60} (1999) 114027;\\
  T.~Nagano, A.~Ota and  Y.~Sumino, Phys.~Rev.\ {\bf D60} (1999) 114014
\bibitem{BogS}N.N.~Bogoliubov and D.V.~Shirkov, ``Quantum fields'',
  Benjamin, 1983
\bibitem{Pivovarov}A.A.~Pivovarov, Phys.~Lett.\ {\bf 236 B} (1990) 214;
  Phys.~Lett.\ {\bf 263 B} (1991) 282
\bibitem{Beneke}V.A.~Smirnov, Phys.~Lett.\ {\bf B404} (1997) 101;\\
  M.~Beneke and V.A.~Smirnov, Nucl.~Phys.\ {\bf B522} (1998) 321;\\
  V.A. Smirnov and E.R. Rakhmetov, Teor.~Mat.~Fiz.\ {\bf 120} (1999) 64
\bibitem{Prudnikov}A.P.~Prudnikov, Yu.A.~Brychkov and O.I.~Marichev,\\
  ``Integrals and Series'', Vol.~2, Gordon and Breach, New York, 1990
\bibitem{ellRef}B.Almgren, Arkiv f\"or Fysik,  {\bf 38} (1967) 161;\\
  S.Bauberger, F.A.~Berends, M.~B\"ohm and M.~Buza,
  Nucl.~Phys.\ {\bf B434} (1995) 383
\bibitem{GKP2}S.~Groote, J.~G.~K\"orner and A.~A.~Pivovarov,
  Phys.~Rev.\ {\bf D60} (1999) 061701
\bibitem{Tokarev}A.A.~Pivovarov and V.F.~Tokarev,
  Yad.~Fiz.\ {\bf 41} (1985) 524
\bibitem{GradshteynRyzhik}I.S.~Gradshteyn and I.M.~Ryzhik,\\
  ``Tables of integrals, series, and products'', Academic Press, 1994
\bibitem{AbramowitzStegun}M.~Abramowitz, I.A.~Stegun (eds.), ``Handbook of
  Mathematical Functions'', Dover Publ.~Inc., New York, 9th Printing, 1970
\end{thebibliography}
\end{document}